\documentclass[conference]{IEEEtran}

\usepackage{amsmath,amssymb,epsfig,cite,footnote,authblk}

\newcommand{\es}[1]{\begin{equation}\begin{split}#1\end{split}\end{equation}}

\newcommand{\R}{\mathbb{R}}

\newcommand{\V}{\mathcal{V}}

\newcommand{\br}{\mathbf{r}}

\newcommand{\dd}{\textrm{d}}

\begin{document}

\title{Connectivity of Cooperative Ad hoc Networks}
\author[1]{Orestis Georgiou}
\author[1]{Georgios Kalogridis}
\author[2]{Hachem Yassine}
\author[1]{Stojan Denic}
\affil[1]{Toshiba Telecommunications Research Laboratory, 32 Queens Square, Bristol, BS1 4ND, UK}
\affil[2]{Department of Engineering Science, University of Oxford, Parks Road, OX1 3PJ, Oxford, UK}

\maketitle


\begin{abstract}
The connectivity properties of ad hoc networks have been extensively studied over the past few years, from local observables, to global network properties.
In this paper we introduce a novel layer of network dynamics which lives and evolves on top of the ad hoc network.
Nodes are assumed selfish and a snow-drift type game is defined dictating the way nodes decide to allocate their cooperative resource efforts towards other nodes in the network. 
The dynamics are strongly coupled with the physical network causing the cooperation network topology to converge towards a stable equilibrium state, a global maximum of the total pay-off.
We study this convergence from a connectivity perspective and analyse the inherent parameter dependence.
Moreover, we show that direct reciprocity can be an efficient incentive to promote cooperation within the network and discuss the analogies between our simple yet tractable framework with D2D proximity based services such as LTE-Direct.
We argue that cooperative network dynamics have many application in ICT, not just ad hoc networks, and similar models as the one described herein can be devised and studied in their own right. 
\end{abstract}

\section{Introduction \label{sec:intro}}

Wireless ad hoc networks are decentralized infrastructure-free networks equipped with multihop relaying and signal processing capabilities and find application in distributed sensor and mobile systems such as smart meters, environmental or industrial monitoring, disaster relief operations etc.
Commonality in many of these applications arises in that the number and distribution of nodes in the networks is often random motivating the study of \textit{random geometric graphs} \cite{penrose2003random}, a statistical framework within which network properties can be modelled, analysed and therefore optimized.
In the simplest case, these graphs consist of a large number of points scattered in a region of space and connected in pairs whenever their separating distance is less than some scalar value $r_0$ as originally proposed by Gilbert in 1961 \cite{gilbert1961random}.
A plethora of generalizations of this basic model have been put forward since then in an attempt to understand the connectivity properties of ad hoc networks and suggest improved network design, routing protocols and deployment methodologies \cite{younis2008strategies}.

Finite, non-convex and complex deployment regions\cite{coon2012full,georgiou2015network,dettmann2014more}, directional and multiple antennas \cite{georgiou2013connectivity,7084129,7239546}, interference effects due to different medium access control (MAC) and trust protocols \cite{baccelli2006aloha,7151120,haenggi2009stochastic,coon2014modelling}, as well as different fading and connectivity models \cite{dettmann2014connectivity} are just a few of the topics that have been studied in recent years.
In most of these works, the main observables of interest relate to pairwise connectivity (e.g.\ outage probability, coverage), the local degree distribution of nodes and multi hop properties (e.g.\ mean degree and hop distribution), and finally more global observables such as the minimum network degree and the full connection probability.
Significantly, these graph connectivity properties have been known to translate to mesh network performance indicators of reliability, routing, end-to-end delay, reachability, interference tolerance, capacity etc., and therefore form the fundamental network skeleton on which higher layer network functions operate.

Peer-to-peer (P2P) is a distributed system architecture designed to share resources such as digital information without the need of a central authority, but rather by direct cooperation between peers \cite{androutsellis2004survey} such as in the case of Gnutella and BitTorrent file sharing networks and more recently the FireChat instant messaging service amongst many others. 
What motivates the use of P2P networks is their ability to function, recover from failures and self-organize in an extremely dynamic and large population of nodes which typically relies on the efficient cooperation between self-interested peers. 
Deciding when and how to cooperate is therefore at the heart of P2P networks and has been extensively researched by Game theory; the study of strategic decision making \cite{myerson2013game} where players (peers) are assumed selfish, and Nash equilibria represent the set of strategies in which no unilateral deviation can make a node extract a higher pay-off.

While the connectivity properties and cooperation dynamics of ad hoc networks have been extensively studied in their respective scientific communities, the interaction between these two networks has not.
In fact, to the best of the authors knowledge most literature focuses on either the connectivity properties or the algorithmic rules which govern network cooperation dynamics.
To this end, in this paper we introduce a novel framework for incentivized cooperation which is intrinsically coupled to the physical network characteristics such as the wireless node locations, statistical channel fading effects of electromagnetic propagation etc.
More specifically, we overlay a Snowdrift game on top of a random geometric graph modelling an ad hoc mesh network consisting of \textit{selfish} nodes and we examine the effects to the cooperation network topology through various connectivity observables.
In other words, we will first model a connectivity network represented through a weighted \textit{connectivity} adjacency matrix $\mathbf{H}$ with constant entries related to the pairwise connection probability (a function of path-loss attenuation and small-scale fading) between nodes, and then model the cooperation network represented through a weighted \textit{cooperation} adjacency matrix $\mathbf{E}$ with entries which evolve according to the Snowdrift optimization dynamics \eqref{adapt}.

Throughout this paper an effort is made to maintain generality in the proposed theoretical framework and avoid specific application areas.
In the conclusion section however we discuss how our proposed framework can be understood within the context of the current LTE-Direct standardization efforts paving the way towards fog networking as the industry progresses into the internet of things (IoT) age and later into the internet of everything (IoE) paradigm.

In this paper we investigate for the first time how self-organizing cooperative relaying protocols affect the network topology and connectivity. 
The main contributions of this paper are as follows: 
\begin{itemize}
\item{we introduce a mathematically tractable and novel framework for network cooperation dynamics based on self-organizing selfish nodes;}
\item{we prove that nodes require a shared benefit for a mixed evolutionarily stable equilibrium state to exist and thus introduce a mechanism for incentivized cooperation to achieve that;}
\item{we describe a numerical procedure for simulating such self-organizing networks and investigate the connectivity properties of the equilibrium state and its dependence on various system parameters.}
\end{itemize}

The remainder of the paper is structured as follows:
Sec.~\ref{sec:model} introduces the system model and also defines a number of network connectivity observables.
Sec.~\ref{sec:dyn} introduces the network dynamics of a Snowdrift cooperation game and also describes how a mutual cooperation benefit between node pairs can incentivize cooperation.
Sec.~\ref{sec:res} presents the methods and results of our numerical investigation regarding the connectivity properties of the evolved steady state of cooperation networks and their dependence on different system parameters.
Finally, Sec.~\ref{sec:conc} summarizes our results and discusses their applicability to ICT applications such as the internet of everything (IoE).


\section{System Model and Connectivity Metrics \label{sec:model}}

In this section we briefly introduce some basic network connectivity observables which will later be used to track and analyse the evolved cooperation network.

\subsection{ Node Deployment}
Consider a two dimensional region $\V \subseteq \R^2$ of area $V$ containing $N$ wireless devices (nodes).  
These are distributed according to an independent Binomial point process (BPP)\footnote{Note that this differs from the usual case of $N \sim \text{Pois}(\rho)$ which results in a Poisson point process (PPP), however for $N$ large enough the two point processes exhibit similar statistical properties and are practically identical.}, with intensity $\rho=N/V$ within $\V$.
Such a configuration is commonly found in WSN applications where sensors or smart meters form a random mesh topology \cite{cordeiro2011ad}.
We denote the locations of the nodes by $\br_i \in \V$ for $i\in[1,N]$ such that the distance between two nodes is given by $r_{ij}=| \br_i - \br_j |$.

\subsection{Path-loss and Fading}
It is known that fundamental results on the connectivity and capacity of dense ad hoc networks strongly depend on the behaviour of the attenuation function \cite{dousse2004connectivity}.
Here, we model the attenuation in the wireless channel as the product of a large-scale path-loss component and a small-scale fading component. 
The former follows from the Friis transmission formula where the long time average signal-to-noise ratio at the receiver (in the absence of interference) decays with distance like $\textrm{SNR}_{ij}\propto r_{ij}^{-\eta}$, where $\eta$ is the path loss exponent usually taken to be $\eta=2$ in free space and $\eta>2$ in cluttered urban environments. 
We therefore define a path-loss function
\es{
g(r_{ij})= \frac{1}{\epsilon+r_{ij}^\eta},
\qquad \epsilon\geq 0,
}
where the $\epsilon$ introduces a kind of guard zone such that for $\epsilon>0$ the path-loss function $g(r_{ij})$ is non-singular at $r_{ij}=0$.
Note that for theoretical analysis purposes $g(r_{ij})$ is unit-less as is $r_{ij}$ which is scaled by the signal wavelength or some other unit of distance.
This has no effect on the results that follow and can easily modified to accommodate for more detailed models.
Finally, for the sake of simplicity and mathematical tractability, the small-scale fading component is assumed Rayleigh such that the channel gain $|h_{ij}|^2$ between two nodes $i$ and $j$ is modelled by an exponential random variable of mean one.
The effects of lognormal shadowing are ignored in our model.

\subsection{Pairwise Connectivity}
Assuming negligible co-channel interference (e.g.\ perfect CDMA/TDMA) and lossless 
antennas, we define the pairwise connectivity through the relation 
\es{
H_{ij} = \mathbb{P}(\textrm{SNR}_{ij} \geq \wp)
\label{H1}}
i.e. the complement of the outage probability, where  the average received signal-to-noise ratio is given by $\textrm{SNR}_{ij}=  g(r_{ij})|h_{ij}|^2 / \beta$, and the parameter $\beta \propto P_T^{-1}$ depends on transmit power $P_T$, the center frequency of the transmission and the power of the long-time average background noise at the receiver ($\beta$ defines the length scale).
We therefore have that $H_{ij}(r_{ij}) = 1- F_{|h_{ij}|^2}(\wp \beta /g(r_{ij}))$,
where $F_{|h_{ij}|^2}$ is the CDF of the channel gain $|h_{ij}|^2$.
Consequently, we obtain
\es{
H_{ij}(r_{ij}) = \exp(-\wp \beta /g(r_{ij}))
\label{H},}
that is, an exponentially decaying with distance pairwise connection probability.
The pairwise connectivity $H_{ij}(r_{ij})$ represents the reliability of the link between nodes $i$ and $j$ but can also be understood as a proxy to the link quality and capacity.
Note that in the limit of $\eta\to \infty$ equation \eqref{H} converges to a step function which for $\epsilon=0$ offers an interesting transition from a \textit{soft} (probabilistic) connection function to a \textit{hard} (deterministic) one corresponding to Gilbert's original model with $\wp \beta=1/r_0^\eta$. Hence $r_0$ gives the characteristic length scale of connection.
Note that we will henceforth assume that the channel is reciprocal i.e. $|h_{ij}|^2=|h_{ji}|^2$.

\subsection{Degree distribution}
The degree distribution is the pdf of the number of 1-hop neighbours.
The probability that node $i$ situated a $\br_i$ connects with some randomly chosen node $j$ can be obtained by averaging over all possible $\br_j$ node positions
\es{
H_i(\br_i)=\frac{1}{V}\int_\V H_{ij}(r_{ij}) \dd \br_j
\label{H2}.}
Since the node positions are chosen according to a BPP, the probability that node $i$ connects with exactly $k$ other nodes (i.e. i has degree $k$) denoted by $d_i(k)$ is given by a binomial distribution which can be approximated by a Poisson distribution for $N,V\gg 1$
\es{
d_i(k)&= {N-1 \choose k} H_i^k (1-H_i)^{N-1-k} \approx \frac{ \lambda_i^k }{k!} e^{-\lambda_i}
,}
where $\lambda_i = (N-1)H_i$. 
Note that when $\V=\R^2$ then the system is homogeneous and we can drop the $i$ index from the above definitions and simply calculate the mean 1-hop degree by substituting \eqref{H} into \eqref{H2} and assuming that $\rho\approx (N-1)/V$
\es{
\lambda = 2  \rho \pi r_0^2 \Gamma(2/\eta)/ \eta = \rho \pi r_0^2 + \mathcal{O}(1/\eta)
.\label{mean}}
The above equation also holds for network 2D domains $\V$ with no borders, i.e. are invariant under translations and rotations e.g.\ the surface area of a sphere, but can also be used as a decent approximation for the mean degree when $N,V\gg 1$.

\subsection{Assortativity}
The assortativity coefficient $r\in[-1,1]$ is a network metric of a node's preference to connect to others that are similar in some way e.g.\ in degree \cite{newman2002assortative} in which case it is defined as the covariance of two random variables $X$ (the degree of a node) and $Y$ (the degree of its neighbours) divided by the product of their standard deviations
\es{
r= \frac{cov(X,Y)}{\sigma_X \sigma_Y}.
}
Algorithms for calculating $r$ are available in most commercial network simulator software packages.
Correlations between node degrees of connected nodes has been observed in many real networks. For instance, in social networks, nodes tend to be connected with other nodes with similar degree values giving $r>0$. 
On the other hand, technological and biological networks are typically disassortative, as high degree nodes tend to attach to low degree nodes giving $r<0$.
Completely random graphs e.g.\ Erd\"os-R\'enyi graphs have $r=0$.
For random \textit{geometric} graphs employing Gilbert's model (i.e. $\eta=\infty$) the expected assortativity coefficient can be explicitly calculated to $r= 1-\frac{3\sqrt{3}}{4\pi} \approx 0.587$ in the high density 
coinciding with the average clustering coefficient \cite{antonioni2012degree}.

\subsection{Weighted Adjacency Matrix}

\begin{figure}[t]
\centering
\includegraphics[scale=0.35]{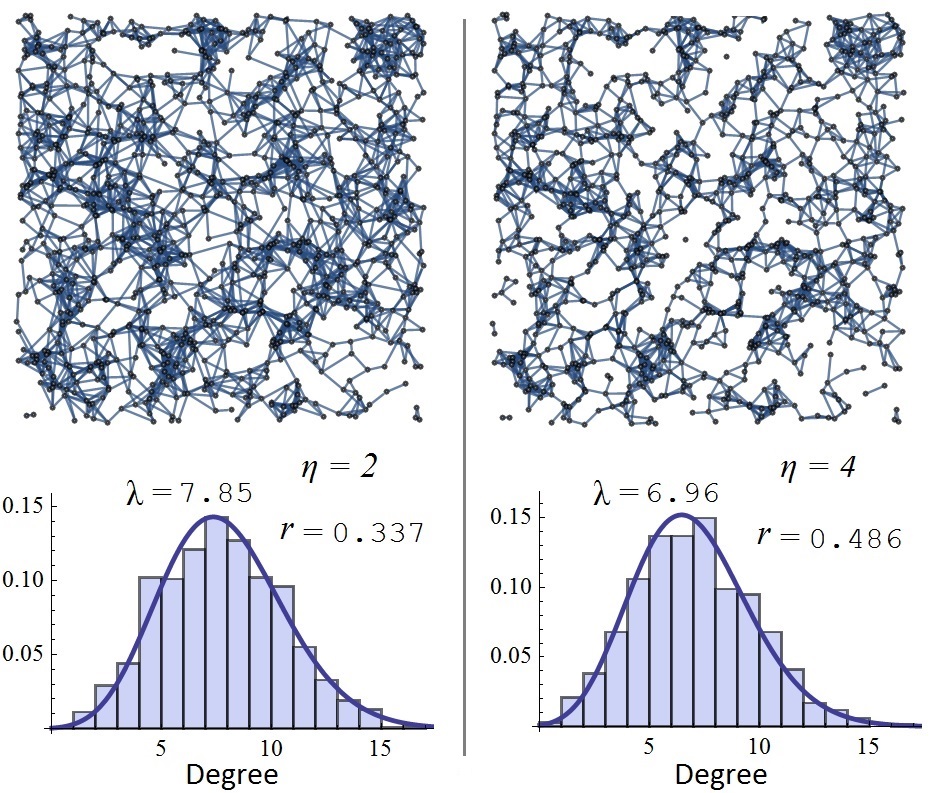}
\caption{Two realizations of a random geometric network using $N=10^4$ nodes, a square domain of side $L=25$, $r_0=1$ and $\epsilon=0$. Both networks use the same node positions however the links are formed at $\eta=2$ (left) and $\eta=4$ (right) path loss exponents. Consequently, these lead to different degree distributions and assortativity coefficients as shown in the lower panels.}
\label{fig:1}
\end{figure}

The adjacency matrix is a means of representing the connectivity of a graph i.e. which nodes are connected to which.
Since the pairwise connectivity as defined by \eqref{H1} is probabilistic, for a given realization of node positions $\br_i \in \V$ we can define a symmetric $N\times N$ weighted adjacency matrix $\mathbf{H}$ with $H_{ij}$ entries and $0$ diagonal.
To realize the edges of this graph, a randomly generated number $\zeta\in[0,1]$ is called $N \choose 2$ times and if  $\zeta\leq H_{ij}$, nodes $i$ and $j$ are paired up.
This guarantees that the network links are statistically independent.
From $\mathbf{H}$, we can therefore calculate all the above mentioned metrics.
For illustration purposes, Fig.~\ref{fig:1} shows two realizations of a random geometric network and its node degree distribution, mean degree $\lambda$, and assortativity coefficients $r$ for different path loss exponents $\eta$.


\section{Network Dynamics and Cooperation \label{sec:dyn}}

We will now introduce a network dynamics in the form of a P2P cooperation network as embodied by a continuous Snowdrift (SD) game.
The SD game is obtained by relaxing the dilemma presented by the celebrated prisoners dilemma (PD)\footnote{The PD embodies the primary problem of cooperation where selfish individuals are better off when exploiting the cooperation of others.
} by assuming that cooperation can result in a shared benefit.
The SD modification is significant, especially in continuous (iterated) games since cooperation is typically maintained through mechanisms of direct and indirect reciprocity, resulting in a mixed evolutionarily stable equilibrium state \cite{doebeli2005models}.
We will show that this shared benefit is essential in incentivizing cooperation and maintaining a connected network.

\subsection{P2P Cooperation Graph}

We construct a cooperation graph through an $N\times N$ weighted \textit{cooperation} adjacency matrix $\mathbf{E}$ with entries $e_{ij}\geq 0$ describing which node collaborates with which and by how much.
For the sake of generality we will not define any particular unit of cooperation or impose any additional constraints on $\mathbf{E}$ other than $e_{ii}= 0 , \, \forall \, i \,$ i.e. self-cooperation is not allowed. 
Note that in general $e_{ij} \not= e_{ji}$.
Every node can cooperate with every other node subject to the connectivity of the graph defined through $\mathbf{H}$ and so the $N(N-1)$ weights $e_{ij}$ are interpreted as a measure of cooperation between node $i$ and $j$ and will be allowed to evolve in time as described in the following subsection.

\subsection{Cooperation Dynamics}

We enable this network to evolve deterministically through the system of $N(N-1)$ partial differential equations describing the selfish nature of peers trying to increase their total pay-off $P_i$ through a downhill gradient optimization given by
\es{
\frac{\dd}{\dd t} e_{ij} = \frac{\partial}{\partial e_{ij}} P_{i} 
\label{adapt}}
where $P_i$ is defined as the total utility pay-off of node $i$ equal to the difference between its total benefits $B_i$ and total costs $C_i$.
What \eqref{adapt} is essentially saying is that the rate of change of the cooperation $e_{ij}$ between $i$ and $j$ is proportional to the variation of $P_i$ with respect to $e_{ij}$.
This reflects the fact that selfish nodes will strengthen/weaken their cooperative links with other nodes in a way that will increase their total pay-off $P_i$.
The evolution of the network stops when $\frac{\partial}{\partial e_{ij}} P_{i} =0$ for all $N(N-1)$ directed edges of the cooperation graph $\mathbf{E}$.
Note that the connectivity network $\mathbf{H}$ remains unchanged throughout.

\subsection{Pay-offs, Costs and Benefits}

In evolutionary biology, costs and benefits are measured in terms of Darwinian fitness (i.e. reproductive success) whilst in other contexts more measurable utility scales are preferable e.g.\ monetary, time, bandwidth, energy, etc. \cite{doebeli2005models}.
In this paper, we will not restrict the discussion to one particular utility metric as that is application specific. 
Instead an effort will be made to keep the introduced cooperation framework application independent.
For the cooperation network described above, the total benefit of node $i$ is without loss of generality some increasing function of the total \textit{incoming} cooperative efforts of peers weighted by the communication link quality of each pair. 
For simplicity we will use a sub-linear function 
\es{
B_i = \sqrt{\sum_{j\not=i} H_{ji} e_{ji}}
,}
such that stronger incoming links (i.e. $H_{ji}\approx 1$) translate into profitable collaborations and \textit{vice versa}.
Similarly, the total cost of node $i$ is some increasing function of the total \textit{outgoing} cooperative efforts that node $i$ is connected to.
For simplicity we will use a super-linear function thus giving 
\es{
C_i = \Big(\sum_{j\not = i } (1-H_{ij}) e_{ij}\Big)^2
,}
such that weaker outgoing links (i.e. $H_{ij}\approx 0$) translate into costly collaborations and \textit{vice versa}.
The sub- and super- linear forms chosen above capture the basic features of real-world systems such as diminishing returns at high cooperation levels as wells as additional costs incurred by the overexertion of ones resources respectively.
Other functions could just as easily have been used in our flexible model.

Therefore the total pay-off of node $i$ is simply given by the difference of the two $P_i=B_i - C_i$ and the cooperation weights $e_{ij}$ are updated according to the $N (N-1)$ system of partial derivatives 
\es{
\frac{\dd}{\dd t} e_{ij} &= \frac{\partial}{\partial e_{ij}} P_{i} = -2(1- H_{ij})\sum_{k\not = i } (1-H_{ik}) e_{ik} \leq 0.
\label{dev}}

\subsection{Tragedy of the Commons \label{sec:tragedy}}

The system described in $\eqref{dev}$ has a clear cooperation problem. Namely we have $\frac{\partial}{\partial e_{ij}} B_{i}=0$ since the benefit $B_i$ of a node depends only on what others are willing to contribute towards it. Therefore, since costs incurred by a node depend on its own contributions we have $\frac{\dd}{\dd t} e_{ij}\leq 0$ for all node pairs.

Significantly, in Gilbert's unit disk model, all not connected pairs will of course not cooperate, whilst all connected pairs will not alter their cooperation weights at all.
For more realistic ``soft" connectivity models where $H_{ij}<1$ (e.g.\ many fading models found in the literature \cite{dettmann2014connectivity}), willingness to cooperate will decay to zero exponentially fast and completely disconnecting the cooperation network i.e. $e_{ij}\to 0$ as $t\to \infty$ for all $i,j\in[1,N]$.
This phenomenon is also referred to as the ``tragedy of commons'' where the presence of cooperators and defectors side by side will cause the extinction of cooperators and the survival of only defectors \cite{doebeli2005models}; not surprising since the proposed model does not provide any incentive to cooperate.

\subsection{A Framework for Insentivized Cooperation}

Direct reciprocity was proposed by R. Trivers as a mechanism for the evolution of cooperation in 1971 \cite{trivers1971evolution}.
We adopt such a mechanism through a term which we call ``a mutual cooperation benefit'' defined as some increasing function of the total \textit{mutual} cooperative efforts per pair that node $i$ is connected to.
In order to capture the inefficiency of small mutual cooperation levels and the saturation of benefits at high levels of mutual cooperation we use a sigmoidal function
\es{
f(x)=\frac{2 b \mu}{\sqrt{\tau+\mu^2}} + \frac{2 b(x-\mu)}{\sqrt{\tau+(x-\mu)^2}}
\label{b}}
where $b=\big(2+\frac{2\mu}{\sqrt{\tau+\mu^2}}\big)^{-1}$ such that $f(0)=0$ and $f(\infty)= 1$ and the positive parameters $\mu$ and $\tau$ specify the inflection point and steepness of $f$ respectively.
Therefore, the total mutual cooperation benefit of node $i$ is given by
\es{
M_i = \sum_{j\not=i} H_{ij} f(e_{ij}+e_{ji})
,}
such that stronger links (i.e. $H_{ij}\approx 1$) translate into stronger collaborations and \textit{vice versa}.
Thus, cooperating results in a benefit which we assume adds linearly to a modified total pay-off $P_i^*$ such that
\es{
P_i^* = B_i - C_i + m M_i
}
with $m\geq 0$ controlling the level of incentivized cooperation.

The inclusion of incentivized cooperation through direct reciprocity changes the dynamics significantly resulting in
\es{
\frac{\dd}{\dd t} e_{ij} =  m H_{ij} f'(e_{ij}+e_{ji}) -2(1- H_{ij})\sum_{k\not = i } (1-H_{ik}) e_{ik},
\label{dev2}}
where $f'(x)$ is the derivative of $f$ with respect to $x$.
Notice that $\frac{\dd}{\dd t} e_{ij}$ unlike \eqref{dev} can now also attain positive values indicating the strengthening of cooperation links.
Moreover, $\frac{\dd}{\dd t} e_{ij}$ only depends on local pairwise properties therefore facilitating real-time P2P networking.

For the sake of clarity, we briefly describe the physical intuition behind the proposed framework captured by \eqref{dev2}:
For $m>0$ the first term on the RHS is non-negative and will therefore attract cooperation towards well connected and cooperative node pairs.
The last term on the RHS is negative and will therefore apply pressure to keep the total cooperation efforts of node $i$ to a minimum especially for badly connected nodes.
Therefore, nodes will shift their cooperation ``investment" weights $e_{ij}$ as to maximize their own total pay-off $P_i^*$.

The $N(N-1)$ system of partial equations described in \eqref{dev2} is a deterministic downhill-gradient optimization problem with no memory (i.e. the state of the system at time $t+\delta t$ only depends on the state of the system at $t$).
Clearly \eqref{dev2} is not separable\footnote{The differential equations cannot be written as a function of a single variable. Instead the time derivative of $e_{ij}$ depends on all $k$ neighbours of $i$.} for $\lambda \gg 1$ and can only be solved through numerical integration.
Moreover, as we shall see in the next section, the numerical exploration of the phase space landscape will converge to a final configuration which is a global maximum of the total pay-off $P^*=\sum_i P_i^*$; a global attractor.
Moreover, and unlike $\eqref{dev}$, the state of equilibrium reached is expected to exhibit some level of cooperation which sensitively depends on the initial topology of the system (i.e. the number and node locations $\br_i\in\V$) and the parameters $(m,\mu,\tau,\epsilon,\eta, r_0)$ used.


\section{Simulation Results \label{sec:res}}

\begin{figure}[t]
\centering
\includegraphics[scale=0.9]{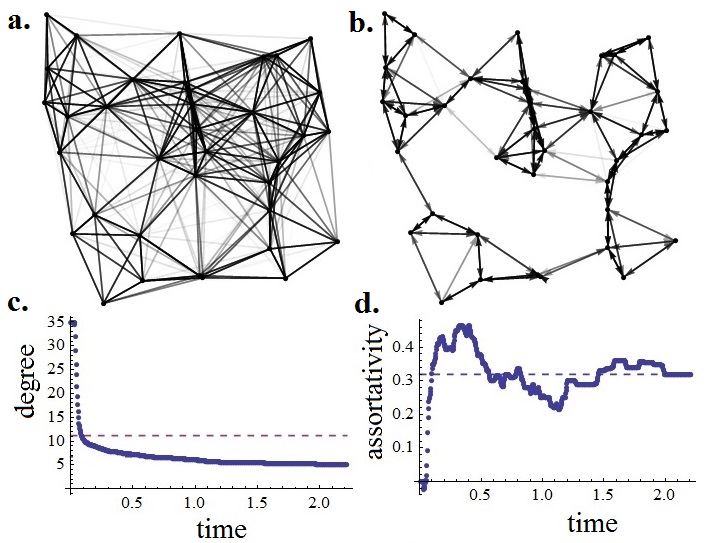}
\caption{\textbf{a.} Connectivity graph with edge weights given by $H_{ij}(r_{ij})$. 
\textbf{b.} Cooperation graph at equilibrium with weights given by $e_{ij}$. 
\textbf{c.} Convergence of the cooperation graph mean node degree $\lambda$ as a function of time. The dashed line is given by  \eqref{mean}. 
\textbf{d.} Convergence of the cooperation graph assortativity $r$ as a function of time. The dashed line is included to guide the eye.
Parameters used: $N=36, \rho=\eta=4, r_0 = \mu = \tau = m=1$, and $\epsilon=0.1 $.}
\label{fig:2}
\end{figure}

We will now simulate \eqref{dev2} and investigate how the connectivity metrics introduced in Sec.~\ref{sec:model} are affected by the proposed framework for cooperation and \textit{vice versa}.

\subsection{Evolving networks}

The time evolution of the P2P cooperation network $\mathbf{E}$ is governed by the downhill gradient given in \eqref{dev2} and strongly depends on the initial state of the system. 
The locations of the nodes form a BPP as described in Sec.~\ref{sec:model}.
For simplicity we choose a uniform initial cooperation state $\mathbf{E}(t=0) = J_N - I_N$ where $J_N$ is a square $N\times N$ matrix with all its entries equal to $1$ and $I_N$ is the identity matrix.
The system of differential equations is then integrated using Euler's method with step size $s=\delta t \ll 1$ according to the update function
\es{
e_{ij}^{t+\delta t} = e_{ij}^{t} + s \Big[& m H_{ij} f'(e_{ij}^t +e_{ji}^t ) 
\\& -2(1- H_{ij})\sum_{k\not = i } (1-H_{ik}) e_{ik}^t  \Big]
.}
If in a given time step $e_{ij}$ becomes negative then $\frac{\dd}{\dd t} e_{ij}$ and $e_{ij}$ are set to zero.
It should be noted that similar integration techniques with variable time steps or higher order correction terms to Euler's method can be more accurate however they do not affect the end result for small enough $s$. 
In our simulations we use $s=10^{-4}$ and assume that the network has converged to its equilibrium state when $|e_{ij}^{t+\delta t} - e_{ij}^{t}| \leq s^2 \,\, \forall \, i\, , j$.

Figure \ref{fig:2} shows an example realization of a cooperation network as prescribed by the above framework.
In the first panel, the connectivity of the cooperation network is shown with the different edges weighted by the pairwise connection function $H_{ij}$.
Since the initial state provided to the evolutionary dynamics has $e_{ij} =1$ for all $i\not=j$ and zero otherwise, then the mean degree of the cooperation graph equals $N-1=35$ i.e. every node starts off at $t=0$ with the option to cooperate with every other node. 
It is clearly seen however in panel c. that the mean cooperation degree decays rapidly until it reaches the mean connectivity degree $\lambda$ as calculated by \eqref{mean}, after which the decay slows down and the cooperation dynamics attempts to reach an equilibrium point which maximizes total pay-off $P^*$.
The convergence towards equilibrium is much slower now as nodes must shift their cooperation weights towards nodes with a mutual and almost reciprocal cooperation weight whilst maintaining total costs relatively low. 
In the process nodes tend to cooperate with similar nodes as captured by the assortativity metric plotted in panel d.
We emphasize that whilst the described dynamic behaviour is specific to the node positions $\br_i \in \V$ and the parameters used, the general picture is actually representative of the proposed cooperation dynamics.
In general, the weighted connectivity graph $\mathbf{H}$ provides a structure for the final cooperation graph $\mathbf{E}(t\to\infty)$, which clearly is a subgraph of the former that manages to preserve the strongest links as can be seen in Figure~\ref{fig:2}.

\subsection{Parameter phase space}

\begin{figure}[t]
\centering
\includegraphics[scale=0.33]{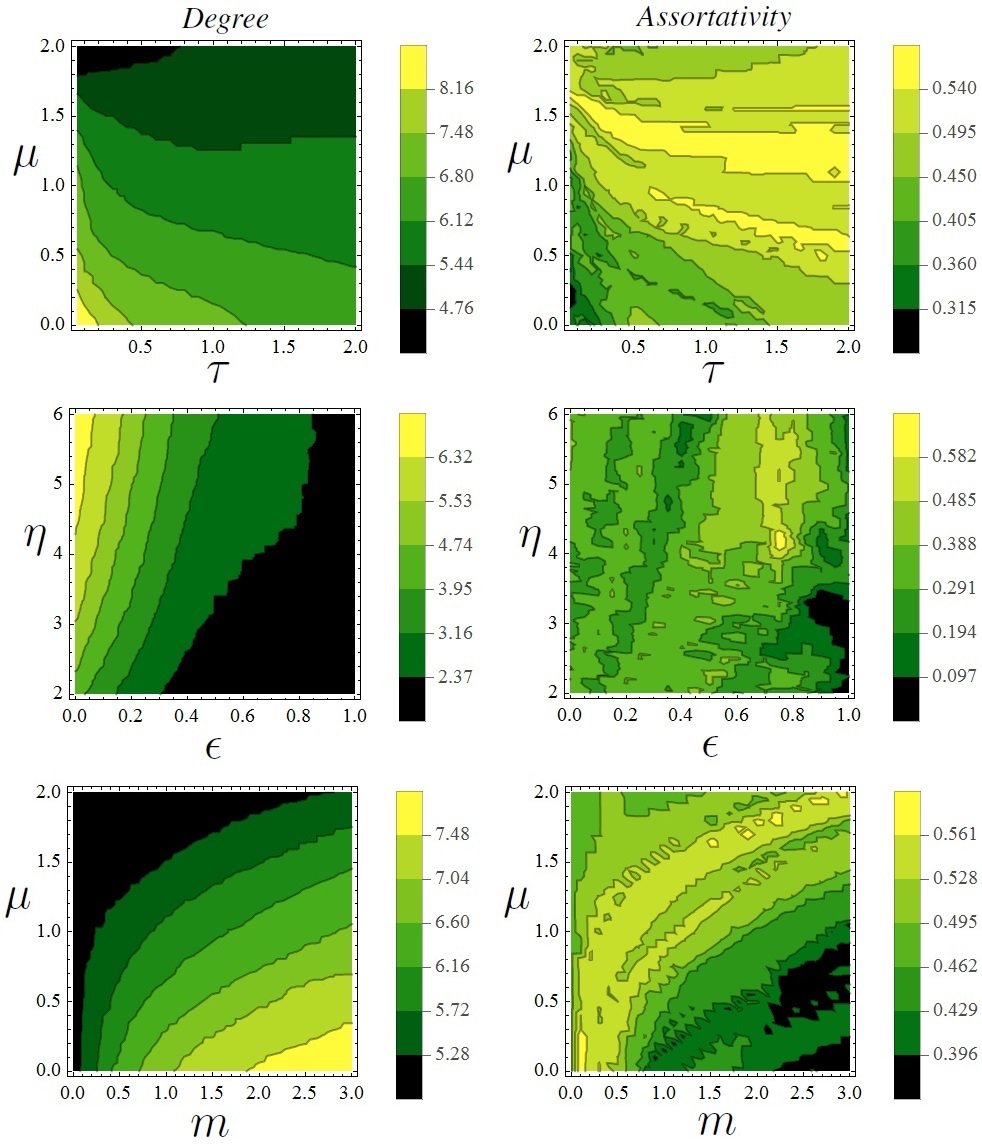}
\caption{
Parameters used: $N=36, \rho=\eta=4, r_0 = \mu = \tau = m=1$, and $\epsilon=0.1 $ unless the parameter is being varied.}
\label{fig:3}
\end{figure}

In this subsection we present further simulation results on the dependence of the cooperation mean degree and assortativity of the evolved graph on system parameters.
Figure~\ref{fig:3} shows that assortativity and degree have a similar dependence to the system parameters which we have investigated.
Indeed the contour plots show similar patterns although assortativity is clearly much more coarse grained than the mean degree.
The reason for this is that associativity is much more sensitive to the network topology than the mean degree and moreover is not a monotonic function of time $t$ as seen in Fig.~\ref{fig:2}d.

\begin{figure}[t]
\centering
\includegraphics[scale=0.33]{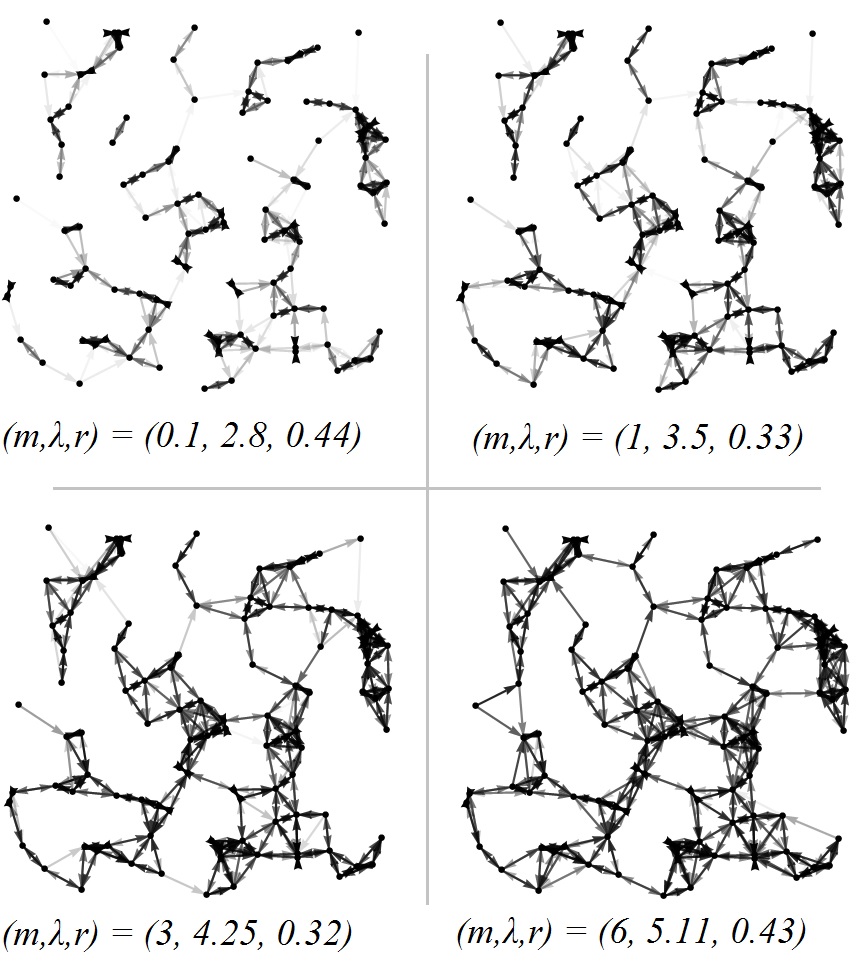}
\caption{
Realizations of the same cooperative network for different levels of incentivized cooperation $m=0.1, 1, 3$ and $6$. 
Parameters used: $N=100, \eta=2, \epsilon=10^{-2}, \rho=2.77, \mu=\tau=r_0=1$.}
\label{fig:4}
\end{figure}

Increasing the inflection point $\mu$ results in a lower degree but a more assortative cooperative graph.
This is expected as cooperation becomes harder to maintain unless nodes are cooperating nodes have are of similar degree.
A similar behaviour is observed when increasing the steepness $\tau$ of the mutual cooperation benefit function $f$.
Increasing the path loss exponent $\eta$ results in a higher degree and more assortative cooperative graph.
This is also expected as a higher $\eta$ means that the pairwise connection function $H_{ij}$ is more step-like and thus well connected nodes $H_{ij}\approx e^{-\epsilon}$ are encouraged to maintain cooperation.
Increasing the guard zone $\epsilon$ has the opposite effect.
Finally, increasing the level of incentivized cooperation $m$ results in a higher degree as expected.
In terms of assortativity however it appears that this initially decreases with $m$ and then increases again. 
We understand this dip in the following way.
When $m=0$, only very well connected connected nodes cooperate whilst the remaining nodes are disconnected. This leads to a high value of $r$.
For small $m\ll 1$, well connected nodes remain cooperative while some not so well connected nodes begin to cooperate, typically with nodes of much higher degree. This increases the mean degree but decreases the assortativity $r$.
For higher values of $m$ the not so well connected nodes strengthen and create even more cooperation links and hence the degree distribution has a smaller variance and assortativity increases.
Figure \ref{fig:4} illustrates precisely this behaviour for a cooperation network at different values of the incentivized cooperation $m$.

\section{Conclusion and Discussion \label{sec:conc}}

In this paper we have studied the connectivity of cooperative ad hoc networks.
The location of nodes are modelled as being randomly distributed in some region and links connecting them are formed according to a path loss function and the fading characteristics of the wireless channel resulting in what is referred to as a random geometric graph.
Nodes are assumed selfish and are incentivized to cooperate with each other  through a snow drift type game which is strongly coupled with the underlying wireless network topology i.e. the quality of the links interconnecting the nodes.
As such each of the $N$ nodes independently decides with which nodes to cooperate and by how much. 
The decisions are such that strongly connected nodes choose to cooperate if the cooperation is reciprocated whilst not to if the link quality is not good enough or if the cooperation is not reciprocated to a satisfactory level.
This novel framework for cooperation is presented through a system of $N(N-1)$ differential equations which is dynamically evolved in time using Euler's method.
Significantly, this paper's focus is on the connectivity of the resulting cooperation graph.
Namely, we have studied the mean degree $\lambda$ and the assortativity $r$ of the resulting cooperation graph and their dependence to various system parameters e.g.\ the path loss exponent characteristic of the wireless propagation medium $\eta$, and the level of incentivized cooperation $m$.

Throughout this paper an effort has been made to maintain generality and avoid specific application areas.
In our opinion the theoretical framework presented could be adapted to a number of ICT problems and presents an interesting new level for research into wireless ad hoc networks.
For example, let us discuss its possible application within the LTE-Direct standardization paradigm, a promising incarnation of device-to-device (D2D) communications based on proximal discovery technologies and LTE licensed wireless spectrum.
LTE-Direct aims to offer an efficient, high-speed method to allow people to connect with and search for people, local businesses and other services nearby through smartphones and other LTE-Direct enabled devices. 
The opportunity to connect directly rather through a base station or access point is clearly a business opportunity, a potential fall-back public safety network, but perhaps also a way to contribute towards the problem of spectrum scarcity, presumably by allowing users to relay data via other D2D devices and onto another LTE-Direct enabled device e.g.\ neighbouring cell base stations or handsets.

The analogies with our (admittedly simple) framework for incentivized cooperation are the following: 
1) Proximity between D2D devices dictates the quality of the wireless link which is inversely proportional with the power cost required by a device to successfully transmit packets.
2) Handset device locations are typically random thus forming an ad hoc network which are often modelled as random geometric graphs.
The most important analogy is however the following:
3) Devices will often have strong wireless links with several other devices, therefore since relaying 3rd party data comes at a personal cost e.g.\ battery consumption and all benefits depend only on the spectrum that other nodes are nodes are willing to contribute, the D2D network is doomed to experience what is known as the tragedy of the commons (c.f. Sec.~\ref{sec:tragedy}). 
An incentive to cooperate is therefore needed. 
This could potentially be resolved through a centralized entity issuing some form of credit, e.g.\ additional data allowance.
This would however defeat the purpose of a distributed D2D system.
A simple alternative to encourage cooperation of devices is direct reciprocity as proposed by R. Trives \cite{trivers1971evolution}, embodied by a mutual cooperation benefit where D2D nodes cooperate with nodes which are themselves cooperative and of course well connected.
Therefore, we propose on top of the proximal discovery a local cooperation algorithm which distributively optimizes a node's willingness to cooperate with nearby nodes based on its available resources and the level of cooperation reciprocated by other LTE-Direct enabled devices.
Indeed, while state of the art ad hoc routing protocols may be used to connect two devices, node benevolence or indirect reciprocity is typically assumed or enforced. 
Instead, the current paper provides an alternative mathematical framework of direct collaboration of selfish nodes that is built upon a wireless connectivity graph and naturally preserves the strongest links while distributively enforcing a sense of mutual cooperation.
The modification of the cooperation framework presented herein to accommodate for node mobility through a time varying connectivity matrix $\mathbf{H}_t$ would be an interesting future research question.

\section*{Acknowledgements}

The authors would like to thank the Directors of the Toshiba Telecommunications Research Laboratory for their support.


\bibliographystyle{ieeetr}
\bibliography{mybib}

\end{document}